\documentclass[11pt,a4paper]{article}
\usepackage{t1enc}
\usepackage[latin1]{inputenc}
\usepackage[english]{babel}
\usepackage{graphicx}

\newcommand{\R}{\mbox{\rm I\hspace{-0.33ex}R}}

\newcommand{\beq}{\begin{equation}}
\newcommand{\eeq}{\end{equation}}
\newcommand{\beqarr}{\begin{eqnarray}}
\newcommand{\eeqarr}{\end{eqnarray}}
\newcommand{\beqa}{\begin{eqnarray*}}
\newcommand{\eeqa}{\end{eqnarray*}}
\newtheorem{theorem(Weyl)}{Theorem (Weyl)}
\newtheorem{lemma(Weyl)}{Lemma (Weyl)}

\begin{document}
\thispagestyle{empty}
\begin{center}
\vspace*{-20mm} 
\section*{An outline of Weyl geometric models in  cosmology} 
{\em Erhard Scholz,\footnote{scholz@math.uni-wuppertal.de}\\
Faculty of Mathematics and Natural Sciences, Dept. of Mathematics\\ 
 Wuppertal University, Gauss-Str. 20\\ 42097 Wuppertal, Germany }  \footnote{Version  March 17, } \\[0.5pt]

\end{center}

\begin{abstract} Here we give a short overview of how  Weyl geometry may be used as a framework for geometrical cosmology.\footnote{An extended version  has been submitted to {\em Annalen derPhysik} \cite{Scholz:2003}.} 
Already the simplest examples of Weyl geometry, the static space-time models of general relativity modified by an additional  time-homogeneous Weylian ``length'' connection (here called the {\em Hubble form}), lead to beautiful cosmological models ({\em Weyl universes}) which can   be considered as a serious alternative to the Friedman-Robertson class presently in use. 
 A comparison with empirical data shows that the magnitude-redshift relation of recent supernovae Ia measurements is in perfect agreement with the prediction of decrease of energy flux in the Weyl models. These data  allow to  estimate the (ex-ante) spacelike curvature $\kappa_0 $ of Weyl universes. The curvature  $\zeta $  measured in cosmic units, i.e.,   $\kappa_0  =\zeta  H_1^{2}$, with Hubble length $H_1 = c^{-1}H_0$, can take on values in a rather wide interval, $ \zeta \in  [-0.4, 2.2 ]$ (for mean quadratic error $\sigma \leq 1.3\, \sigma_{data}$).  Quasar frequency data from the SDSS  provide strong evidence  of a positive ex-ante curvature $\kappa_0 $  and  reduce the interval for  $ \zeta$ to $ \in  [0.25, 2.2 ]$ (for $\sigma \leq 1.3\, \sigma _{best}$). Thus an {\em Einstein-Weyl universe } (i.e., an Einstein universe endowed with a  Weylian length connection) is  in  good agreement with supernovae and quasar data. 

The relative mass-energy density $\Omega_m$, with respect to the critical density of the standard approach,  and the relative contribution of  the ``vacuum term'' $ \Omega_{\Lambda} $ (where  $\Lambda$ denotes  the cosmological term of Weyl geometry) are here identical to $\Omega_m= \zeta$  and $\Omega_{\Lambda} = 1$. Both are time-independent. Thus the time-evolution anomaly of vacuum energy does not arise. The intervals given in the literature for the dynamically determined matter density  parameter $ 0.1 \leq \Omega_m \leq 0.5$ or $0.2 \leq \Omega_m \leq 1$, depending on scales of stucture observed, with presently preferred value $\Omega_m \approx  0.3$ are consistent  with other data in the Weyl geometric approach, although they may not be the last word. An explanation of the cosmic microwave background by a maximal entropy state of the quantized Maxwell equation, according to a proposal by I.E. Segal for the conformal Einstein universe, translates directly to the Weyl approach. It  leads to quantitative estimations for the  temperature of the Planck radiation between 2°K and 5.3°K. Moreover,  the anisotropy angles, expected to arise   in  an Einstein-Weyl universe of $\zeta \approx  0.8 $   from  the  inhomogeneities in matter distribution on the supercluster level give a value of $ l \approx 200$ and agree  with the results of multipole measurements of CMB anisotropies (similarly  in the case of $\zeta \approx  0.3$ for cellular structures). 

\end{abstract}
Keywords: Weyl geometry, Einstein-Weyl universe, cosmological redshift \\[0.5pt]
PACS numbers: 02.40.Ky, 04.90.+e, 98.80.Jk
 \vspace{-7mm}\\
\begin{small}
\end{small}

\subsubsection*{Basics of Weyl universes }
Weyl universes are particularly simple manifolds $M$ with a Weylian metric $(g, \varphi)$. 
Pragmatically speaking, a {\em Weyl universe} arises from a modification of a   Lorentzian product  manifold $M$,  
\[M = \R \times S_{\kappa_0} ,\]
by a Weylian length connection $\varphi$.
Here  $S_{\kappa_0}$  represents  spacelike fibres and denotes a Riemannian 3-manifold of constant ex-ante sectional curvature $\kappa_0 = \frac{k}{a^2}$;   $k \in {0, \pm 1}$ characterizes  the curvature type (Euclidean, spherical, hyperbolic) and $a$ is the curvature radius.
Let us denote  the  3-metric of the fibres by $d\sigma ^2 $.  For coordinates $(t, x) \in M$, the Lorentzian metric on $M$ is then given in geometrical units ($c = 1$)  by
\beq g: ds^2 =   dt^2 - d\sigma ^2 .  \eeq
It is complemented by a differential one form, the {\em Hubble form}
\beq  \varphi = H dt ,\eeq
with $H \in \R^+$ a constant, the {\em Hubble constant}. In physical units its inverse $H_0^{-1}$ represents the Hubble time and $H_1^{-1} = c H_0^{-1}$ the Hubble length. The pair $(g, \varphi)$ characterizes the {\em Weylian metric}, although it is only one of several different {\em gauges} \cite{Weyl:InfGeo}. The Hubble form defines a scale (length) connection  on $M$.  A change of gauge from $(g, \varphi )$ to $(\tilde{g}, \tilde{\varphi })$ is given by a strictly positive scaling function $\lambda : M \rightarrow \R^+$, such that
\[\tilde{g} =  \lambda^2 g \, , \;\;\; \mbox{and} \;\;\; \tilde{\varphi} = \varphi - d \log \lambda . \]
As the Hubble connection has a vanishing  curvature,
\[  d \varphi = 0 , \]
the length connection can be integrated and leads to a semi-Riemannian manifold structure on $M$. After reparametrization of the time coordinate $t = \log \tau$, the semi-Riemannian structure on $M$ can be percieved as a Robertson-Walker manifold  $\R \times _f S_{\kappa_0}$ with  linear warp function $f( \tau ) = H \tau$. In this sense, Weyl universes are nothing but uncommon, but not at all  useless, views on the simplest class of non-trivial Robertson-Walker manifolds. 

The cosmic flow $U$ of the corresponding cosmological model is given by the canonical flow along  $t$, lifted to $M$, i.e., $U = \frac{\partial }{\partial t}$ in Weyl gauge. 

The Weylian metric has a uniquely determined compatible affine connection $\Gamma$ with a gauge invariant Riemannian curvature tensor $R$ and Ricci tensor $Ric$. While $\Gamma$,  $R$ (of type $1,3$) and $Ric$ are gauge invariant,  the scalar curvature $S$ is gauge dependent with gauge weight -2. In our gauge it is constant. Therefore our reference gauge given by equations (1) and  (2)  is called the {\em Weyl gauge} of the manifold.
 If  we express the  sectional  curvature of $S_{\kappa_0}$ by a  proportionality factor $\zeta$ to the  ``Hubble curvature''  $H^{2}$,  i.e. $\kappa_0 := \zeta{H^{2}}$, the scalar curvature of $M$ becomes
\beq \label{scalar}  S = - 6 H ^2 ( 1 + \zeta) . \eeq
The effective spacelike sectional curvature $ \kappa_S$ of the Weylian manifold $M$  (i.e. including the effect of the scale connection $\varphi$) is
\beq \label{sectional}  \kappa_S = H^2  +  \kappa_0    .\eeq

If we use the natural convention that the gravitational constant $G$ is a gauge quantity of weight 2, the Einstein equation
\beq  \label{E-e} Ric - \frac{1}{2} Sg = 8 \pi \frac{G}{c^4}T \eeq 
becomes  gauge invariant. Because of  (\ref{scalar}),
 the l.h.s. of the Einstein equation is in our case
\beq \label{l-h-s} Ric  - \frac{1}{2} Sg =  3 (H^2 + \kappa_0) dt^2   -  (-1) (H^2 + \kappa_0) d\sigma^2 .  \eeq

Thus the energy-momentum-stress tensor $T$ of the  Einstein equation is a tensor of an ideal fluid, although a  ``strange'' one (negative pressure),  with total energy-mass density give in geometrical units ($c=G=1$) 
\beq \label{rho}  8 \pi  \rho =3 (H^2 + \kappa_0)  \eeq
and  pressure 
\[ 8 \pi p = -(H^2 + \kappa_0)  .\]
This is a direct consequence of general results on Robertson-Walker manifolds, considered in Weyl gauge \cite{ONeill}.\footnote{ ONeill works with signature (3,1) of the Lorentz metric rather than (1,3), corresponding to a  factor $ -1$ for the metric. The consequences for $S$, $R$, $Ric$ can be read off from the corresponding gauge weights ($ \lambda \equiv  i$).}
 It is  important to notice that both values are {\em constant in Weyl gauge}.

The constancy of energy density explains why  we consider Weyl gauge   as  an appropriate  candidate for the  representation of measurement by  material (atomic) clocks. From the point of view of the  physical interpretation (``semantics'') of the model,  Weyl gauge $(g, \varphi)$ ought to be considered  as the  {\em matter gauge} of a Weyl universe.  The  accompanying Hubble form represents the scale connection of the Weylian metric; it encodes the cosmological redshift completely.

It is possible to integrate the scale connection of a Weyl universe and to ``gauge it away''. Then we arrive at the {\em Riemann gauge} of the Weylian manifold which has, by definition, a vanishing scale connection $\tilde{ \varphi } =0$. It  is  of the form $(\tilde{g}, 0)$ with $\tilde{g}(t, x) = \lambda (t)^2 g (t,x)$, where the scaling function $\lambda$ arises from (path-indedent) integration of the Hubble form from some reference time $t_0$ (``today'') to any $t$:
\beq \label{Weyl-to-Riemann-gauge}  \lambda (t) = e^{H (t_0 + t)}.  \eeq
Thus   the scaling function from Weyl to Riemann gauge  expands  exponentially.  It can be  reparametrized by a  new timelike parameter $\tau$ in $\R^+$ with
\[  \tau = H^{-1} e^{Ht}, \;\; t = H^{-1} \ln H \tau . \]
Then $M$ gets the form of a  Robertson-Walker manifold, 
\[ M \approx  \R^+  \times_f S_{k,a} \]
with a linear warp (``expansion'') function $f(\tau ) = H \tau$.
In the Robertson-Walker picture of the Riemann gauge, the scale expansion  looks no longer exponential but linear (due to reparametrization). An attempt to    extend the time-like interval from $\R^+$ to $ \R^+ \cup \{ 0 \}$ runs into an ``inital singularity'' as an obstacle.  Clearly the Robertson-Walker picture  suggests an inappropriate method of past timelike compactification of the Weyl universe. 

In Weyl geometry it is natural to describe the  transfer of scaled quantities (metric, time intervals, photon energy, flux of a cosmic source, etc.) by a {\em calibration by transfer} as it was called by H. Weyl, i.e., transfer by integration of the length connection, exponentiated according to the gauge weight of the quantity. In our simple case, the transfer is given path-independently by  $\lambda $ as in equation (\ref{Weyl-to-Riemann-gauge}).
The redshift of a photon with trajectory $\gamma $, emitted  at $p = (t, x)$ and observed at $q = (t_0, x_0)$ ( $t< t_0$ = ``today'' ), is given by
 \[ z (p,q) +1 =  e^{\int _t ^{t_0} \varphi (\gamma '(\tau )) d \tau } = e^{H (t_0- t)} .\]
 For $t_0 = 0$ and $t < 0$ it can be written as
\beq \label{z}  z(t) = e^{H |t|} -1 .\eeq
That agrees, of course, with the value expected from the Robertson-Walker picture. In Riemann gauge and the Robertson-Walker picture Weyl universes no longer  have  a constant energy-momentum tensor. In this sense,  Riemann gauge leads to an {\em unphysical appearance} of the model and is {\em no appropriate candidate for the matter (atomic clock) gauge}. 

 Photon energy (gauge weight -1) is dual to  oscillation time  (gauge weight +1); their   calibration by transfer  is closely linked to Riemann gauge (weight of metric +2). Therefore we  consider the Riemann gauge of Weyl universes as their {\em photon gauge}. Flux calibrates differently (gauge weight  -4). 

The decrease of energy flux  from a cosmic source with redshift $F(z)$ can  now  be calculated  and, with it, the decrease of luminosity $m(z)$ of cosmic sources. If we take the correction of measured data  by a factor $z+1$ into account, usually applied by astronomers in order to compensate for the energy decrease of photons by the redshift ($K$-correction), the luminosity function is given in the spherical case, $k= +1$ by 
\[ m_{+} = 5 \log \left(   (z+1)^{\frac{3}{2}} \sin( \sqrt{\zeta }   \ln (z+1) ) \right) + C .  \]
In the hyperbolic case ($k = -1$)  the $\sin$ has to be substituted by the $\sinh$ and in the ``flat'' case ($k=0$) by the identity function.

The lightcone of a Weyl universe intersects the spatial fibres in  Euclidean, ``spherical'' or hyperbolic spheres, depending on the value of $k$. The dependence of the area of the spheres $O(z)$ with  redshift  $z(t)$  can easily be calculated, and with it the volume  $V(z_1,z_2)$, covered by the lightcone between redshifts $z_1$ and $z_2$.    For the case of an {\em Einstein-Weyl universe}, i.e., the spherical case with $k=1$, we get
\[  {O}_+(z)  =    \frac{ 4 \pi}{ \zeta H_1^2 }   (z+1)^{-2} \sin ^2 (  \sqrt{\zeta }   \ln  (z+1) )\]
and 
\beq \label{volume-increase} V(z_1, z_2) = H_1^{-1} \int_{z_1}^{z_2} (z+1) ^{-1} O (z) dz ,  \eeq
respectively (in physical units).

In an Einstein-Weyl universe of (ex-ante) curvature radius of spatial fibres $a = (\sqrt{\zeta }  H_1)^{-1}$, an object of diameter $b$ in distance $d $ in Weyl gauge ($ d = H_1^{-1} \ln (z+1)$) is  given by calculation in spherical trigonometry, because of the conformal invariance of the lightcone structure (during the deformation of the classical Einstein universe to the Einstein-Weyl one). Angular sizes of such objects  are given by 
\beq \label{angular-size} \sin \frac{ \alpha  }{2 } = \frac{ \sin\frac{ b }{2 a } }{\sin \frac{ d }{ a} } .\eeq

Two Weyl universes with parameters $H$, $H'$ and $\kappa_0 = \zeta H^2$, $\kappa_0' = \zeta ' H'^2$ are isometric iff $\zeta '= \zeta $. Thus, although in the definition of Weyl universes the two parameters $H$ and $\kappa_0$ enter, the model space is essentially 1-dimensional. It can be parametrized by the curvature index $\zeta$. For any empirical investigation we we should therefore be interested in finding approximations of the metrical parameter $\zeta$.

\subsubsection*{Luminosity-redshift relation for supernovae Ia}

Let aus compare  the ``prediction'' of luminosity decrease  in Weyl universes  with the recent supernovae Ia data,   as measured in   highest  possible precision with  present techniques by the {\em Cerro Tololo Inter-American Observatory} group (CTO) around S. Philipps  and  the {\em Supernova Cosmology Project} (SCP) around S. Perlmutter. Integrated results have been presented in \cite{Perlmutter:SNIa}. The whole set of data covers the measurements for 60 supernovae Ia (SNIa) with redshifts  $0.014 \leq z \leq  0.83$ and magnitudes  $14.47 \leq m \leq 24.65$ in two   clearly separated subsamples, 18 low redshift  supernovae observed by the  CTO group with $0.014\leq z \leq 0.101 $, and 42 by the SCP  with $0.172 \leq z \leq 0.83$. The error intervals given for the measurements carry  a mean quadratic error $\sigma _{data} \approx 0.256$.
A  fit by the method of quadratic errors $\sigma$  (residual dispersion)  shows good results for the Weyl model.

\unitlength1cm
\begin{figure}[bottom]
\center{\includegraphics*[height=5.5cm, width=9.0cm]{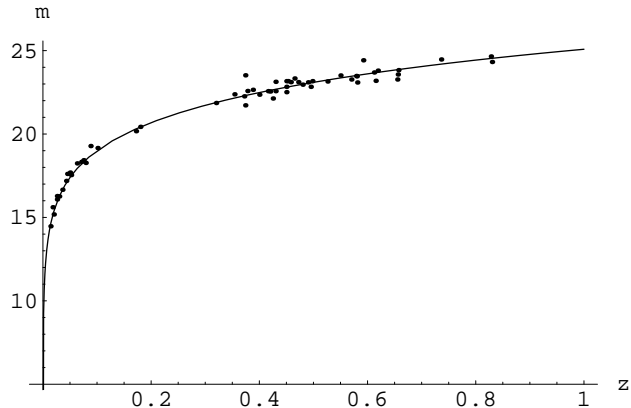}}
\caption{Magnitude-redshift relation in  Einstein-Weyl universe, $\zeta = 1$ (i.e., $\kappa _0 = H_1^2$), $H_0 = 72$, $M= - 19.3$  }
\end{figure}

We use the present value for the Hubble constant $H_0 = 72\; \pm 5 \;km s^{-1} Mpc^{-1}$, determined  by other methods and accepted in the astronomical community (with only minor differences left, we calculate here with  $H_0 \approx  72$) and   $M \approx -19.3$ for the absolute magnitude of supernovae Ia.
With  these values, the best fit for the curvature parameter is 
\[  \zeta \approx 0.9 \, , \; \;\;  \mbox{ with residual dispersion} \;\;  \sigma_{best} \approx  0.324 \approx 1.27 \sigma _{data}. \]
This fit is reasonably  good; the quality is comparable to  the standard model ($\sigma_{stand} \approx  0.322$).\footnote{The value given here  differs  from the one of the extended preprint version  \cite{Scholz:2003} (corresponding to curvature parameter $\zeta \approx 2.1$), due to a correction of the calculation.}
 It can be considered as a first  empirical answer to the question, whether the hypothesis of luminosity decrease according to calibration by transfer in Weyl geometry makes sense or not. It clearly does.
Moreover, the decrease  of fit quality under change of  $\zeta$ is rather slow. With a  criterion  $\sigma \leq 1.3 \, \sigma_{data}$ for a {\em good fit}  we get a wide interval for admissible curvature parameters, including all three curvature types $k= \pm 1, 0$
\[  \zeta \in [ -0.4, 2.2 ]=: I_{SNIa} .\]
The residual dispersions for $\zeta = 0.9$ and $\zeta =0.3$ are $\sigma _1\approx 0.324$ and  $\sigma _{0.3} \approx 0.326$, respectively. The last value ($ \zeta \approx 0.3$)  is of interest because of the considerations below (mass density).

\subsubsection*{\bf Quasar frequencies }
Quasar counts  depending on the redshift and their relative frequencies  in equal $\Delta z$ intervals have shown  intriguing properties, not easily to be explained by  geometrical considerations in the known cosmological approaches. We consider here the  data from the first data release of the {\em Sloan Digital Sly Survey} (SDSS) with 16 713 objects up to $z \leq 5.4$, which  has  been   published recently  \cite{SDSS:2003b}.  The data of the quasar count have been concentrated  in a histogram with  69 intervals of width $\Delta z = 0.076$. They show a clear and well-formed increase up to about $z\approx 2$ and a rapid break down for $z > 2$.

If we compare the data of quasar distribution with the volume scanned by the lightcone in an Einstein-Weyl universe  over corresponding redshift intervals, we find a surprisingly good agreement up to the peak at $z \approx 2$ (see figure 2). We calculate how many objects are to be expected in the redshift intervals of lengths $\Delta z = 0.076$  up to $z_{cut} \leq 2.05$ {\em under the assumption of equal distribution per unit volume} for different values of $\zeta $ in the interval determined by the supernovae Ia data. If the cut-off value is chosen above 2.05, the  fit optimum is influenced by the the rapid decrease of quasar counts above $z \approx 2$.\footnote{The cut-off value $z_{cut} = 2.05$ is specified  by comparison with the fit behaviour of other cut-off values. For lower values (of the cut-off) than $2.05$ the best fit parameter remains essentiall stable $\zeta_{best} \approx 0.8$. For values  $z_{cut} > 2.05$ the upper flank of the rapid decrease of quasar frequencies leads to the appearance of another (and stronger) relative optimum $\zeta_{best}$ above 2, which shifts with changing values for  $z_{cut}$. The latter has been discussed in the preprint  \cite{Scholz:2003}.}
If we  assume that the rapid decrease of quasar counts beyond $ z \approx 2$ is mainly due to a reduction of sensibility of detectors and other selection effects, it seems appropriate to reduce the fit interval as above. 
The result is a best fit value for $\zeta \approx 0.8$. If we use the  criterion $\sigma_{\zeta} \leq 1.3 \, \sigma_{best}$, the  quasar frequencies lead to an admissible   parameter  interval 
\[  \zeta \in [0.25, 2.2]  =:I_{QF} .\]

\begin{figure}[h]
\center{\includegraphics*[height=7.5cm, width= 10cm]{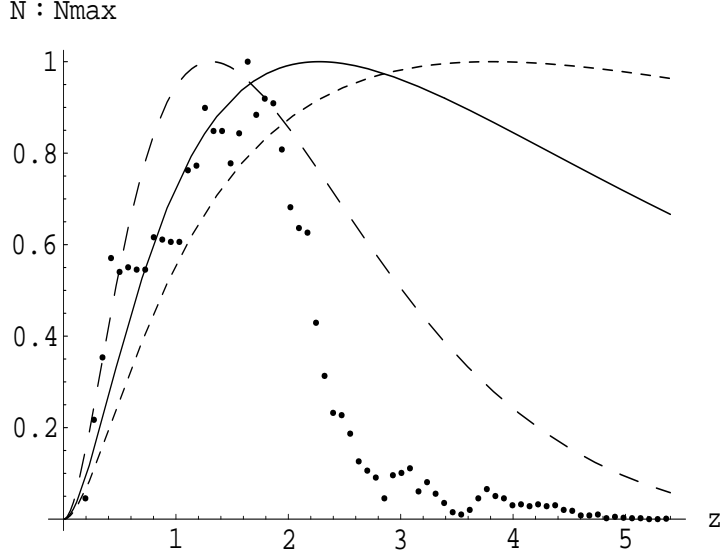}}
\caption{Volume covered by the lightcone in Einstein-Weyl universe with $\zeta =0.8$ (undashed), $\zeta =2.2$ (coarse dashing),  and $\zeta =0.25$ (fine dashing), compared with relative numbers of quasar frequencies of SDSS (dots)}
\end{figure}

If astronomers find such strong selection effects unlikely, the assumption of equal distribution of quasars in sufficiently large  volume layers of the lightcone in Weyl geometry  would be refuted for $z> 2$. 

In any case, the quasar frequencies seem to  give strong evidence in favour  of a {\em positive curvature } parameter if we interpret the data in the frame of the most simple models of  Weyl geometry (Weyl universes). The quasar frequency interval $I_{QF}$ is contained in  the good fit interval for the SNIa data, and we find as common result
\[ \zeta  \in I_{SNIa} \cap{QF} = I_{QF} = [0.25, 2.2]. \]

\subsection*{Weylian cosmological term and  mass density  }
In case of a non-vanishing length curvature $\varphi$, the l.h.s. of equation (\ref{E-e}), i.e., the gauge invariant Einstein tensor,  consists of a part derived from the semi-Riemannian component $g$ of the complete Weyl metric $(g, \varphi)$ and a contribution from the Hubble connection. Let us denote  Levi-Civita connection, Ricci curvature and scalar curvature of the semi-Riemannian component alone by $\Gamma_g, \, Ric_g,\, S_g$ respectively; then the total (gauge invariant) Einstein tensor splits up into two terms
\beq  Ric - \frac{1}{2} S g = Ric_g - \frac{1}{2} S_g g  + \Lambda_H ,\eeq
where  $\Lambda_H$ can be defined by this equation (of course, the split is {\em not} gauge invariant, but depends on the choice of the matter gauge of the model). It expresses the deviation of the gauge invariant Einstein tensor from the underlying purely semi-Riemannian component in the chosen gauge. Thus $\Lambda_H$ can be read as a ``cosmological term'' of Weyl geometry; we  call it the {\em Weylian cosmological term}. In analogy to   standard cosmology,  it seems natural to decompose the r.h.s of the Einstein equation accordingly. Written in geometrical units, we get  
\[ 8 \pi T_{tot} = 8 \pi (T_m + T_{\Lambda} ), \]
where we denote the total energy-momentum tensor by $T_{tot}$ and decompose it (in geometrical units)  according to   equation (\ref{l-h-s}):
\beqa    8 \pi T_m &=& 3 \kappa _0 dt^2 - (-1)\kappa _0 d\sigma ^2 =Ric_g - \frac{1}{2} S_g g \;  \\
  8 \pi T_{\Lambda} &=&
 3 H^2 dt^2 - (-1) d\sigma ^2 = \Lambda_H  \; .\eeqa


This decomposition shows a  natural correspondence between the sectional curvature and mass-energy density on the on hand side, and the curvature contribution of the Hubble form and ``vacuum'' term on the other  (\ref{sectional}), (\ref{E-e})):
\[ 8 \pi \rho_{tot} = 8 \pi (\rho_m + \rho_{\Lambda})  = 3 \kappa_S = 3 \kappa_0 +  3 H^2 . \]
In the light of the established interpretations of standard cosmology, it {\em may} therefore seem  natural to interpret  the curvature values by 
\beqarr \label{split}
 8 \pi \rho_m &=& 3 \kappa_0 \;,  \;\;\; \;  8 \pi p_m =  - \kappa_0     , \\
  8 \pi  \rho_{\Lambda} &=& 3 H^2  \; , \;\;\;\;  8 \pi p_{\Lambda} = - H^2 
\eeqarr 
(in geometrical units). The ``vacuum'' term represents the influence of the Hubble form on the total curvature, while the mass-term seems to curve  the spatial fibres ``directly''.

For the relative contributions compared  with the critical  mass-energy density of the standard approach, $\rho_{crit} = \frac{3}{8 \pi} H^2$, we  get
\[  \Omega_m := \frac{ \rho_m}{\rho_{crit}} = \frac{3 \kappa _0}{3 H^2} = \zeta  \;, \;\;\;\ \Omega_{\Lambda} := \frac{ \rho_{\Lambda}}{\rho_{crit}} = \frac{3H^2}{3 H^2} = 1 .\]

The determination of dynamical mass density, close to galaxies, clusters, and superclusters leads astronomers and physicists to an actual estimation of 
$ 0.1 \leq \Omega_m \leq 0.5  $ for smaller structures and/or $ 0.2 \leq \Omega_m  \leq 1 $ for larger ones,  with a preferred value close to  $ \Omega_m \approx  0.3 $ \cite{Carroll:Constant}. 

That is consistent with the supernovae and quasar data, but probably not the best we can expect. As we expect a greater awareness for dynamical mass in larger regions in the future, we choose the cautious estimation for large structures
$ I_m :=  [\, 0.2, 1 \, ]  $. That is in very good agreement with $I_{SNIa} \cap I_{QF}$, if we adapt the lower bound slightly. A provisional result of all the three aspect considered  is then:
\[    \zeta  \in [\, 0.25 , 1 \, ] =  I_m \cap I_{QF} \cap I_{SNIa}.\]

There are reasons to expect further changes for the estimations of $\Omega_m$.
The dynamically determined mass may be the result of diluted baryonic matter, mostly ionized hydrogen (the {\em hyle} in our terminology), which is  concentrated in the neighbourhood of clusters and superclusters, but not confined to it. Apparently the hyle is  distributed in high dilution  over the whole time-space, including  the voids, in addition  to its dynamically observable more concentrated parts. We should  not exclude the possibility that future investigations may lead to estimations of the overall mass density,  which correct the actual values of $\Omega_m  \approx 0.3$ by a factor 2 to 10. If the higher factor turns out to be realistic, the split of the source quantities of the Einstein equation  (\ref{split}) is  a purely formal exercise which elucidates the analogy to present standard cosmological argumentation, but has no deeper  physical meaning. Please notice that in the Weyl approach $T_m$ and $T_{\Lambda }$ are strictly proportional and obey the same state equation
\[  p = - \frac{ 1 }{3 } \rho \;  .  \]
   From the mathematical point of view, a joint semantical interpretation of $T$, without separating it into its part, seems therefore just as  natural as the decomposition discussed above.

The present estimations of dynamical mass  prefer a value $ 0.25 \leq \zeta \leq 0.3 $, while  the evaluation of supernovae Ia data and of quasar frequencies  indicate  a value of $\zeta $ closer to 1 as a realistic candidate. For conservative estimations  we    therefore  have to calculate  with $\zeta \approx 0.3$; for  general  considerations we give  $\zeta  \approx  0.8$ a slight preference.

\subsection*{Cosmic microwave background  }
I.E. Segal has   proved an interesting property of the {\em conformal} Einstein universe: According to his analyis, the quantized Maxwell field on the  Einstein universe  possesses  a maximal entropy state  of  {\em exact Planck characteristic} with energy distribution $\epsilon (\nu)$ ($\nu =$  frequency)
 \[  \epsilon  (\nu ) = \nu ^3 (e^{\beta  \nu } - 1), \;\; \beta  = 2 \pi \hbar \left(  \frac{\pi ^2}{15 (\hbar c)^3 \epsilon _0} \right)^\frac{1}{4}, \]
where $\epsilon _0$ is the energy density of the radiation field \cite{Segal:CMB1}. 
If we accept this theorem, the {\em theoretical existence of a background equlibrium of the quantized Maxwell  field in a  spherical Weyl universe has   been established by  I.E. Segal}, because electrodynamics is a conformally invariant theory and the Einstein-Weyl universe is a conformal deformation of the classical Einstein universe.

If the Weyl geometric approach is not only a striking formal model of cosmic geometry but has  physical content, we may reasonably expect that the Hubble connection $\varphi = H dt$ is a classical expression of some energy transfer process in the  electrodynamical field from well directed ``foreground contributions'' of the field (well modelled by the classical Maxwell equation) to the quantum electrodynamical ``background'',  which  we consider to be modelled by Segal's maximal entropy state. In this sense, the Hubble connection would  represent a variant of the classical Olbers effect of  background galaxy radiation. In the Weyl geometric context it is delivered, however, to the background rather than to the visible ``night sky''. The  observable background microwave radiation  results only    as a secondary effect. This  background radiation, considered on the classical level, will be exposed to absorption. 

We thus  expect an energy equilibrium in Segal's background state  between transfer processes from the foreground radiation via the Hubble connection (Olbers-Weyl effect) and an energy loss by absorptive damping. We call this hypothetical mechanism the 
{\em Segal-Olbers-Weyl effect} and show that it gives an alternative explanation of the cosmic microwave background radiationt in the frame of Weyl geometry.

We estimate the  absorptive damping   during the passage in cosmic time-space between two conjugate points of the spherical space section by a factor $\alpha $ lying in  a rather genereous interval,   $0.11 \leq \alpha \leq  0.99$. Then   the equilibrium between absorption and energy  transfer  from  the foreground galaxy radiation field, with energy density  $ \epsilon _{gal}$,  is obtained at 
\[ \epsilon _{back} = \alpha ^{-1} \epsilon _{gal}\frac{z_{con}-1}{z_{con}} , \]
where $z_{con}$ denotes the redshift at the conjugate point of the observer. It can be used for an estimation of the background energy density,

Recent measurements of the cosmic infrared, optical and UV extragalacitc radiation   give values for its overall intensity,  $ I_{IOUV}$, between 45 and 170 $nW\, m^{-2} sr^{-1}$ \cite[41]{Dwek/Hauser:2001}. That corresponds to an energy density  $\epsilon _{IOUV}$ between  $0.012$ and $0.045$  $eV cm^{-3}$ (ibid, p. 3). Thus the energy density has  a  proportional  (``geometric'')  mean 0.024 $eV cm^{-3}$  with error factor $2^{\pm 1}$. For the absorption interval  $\alpha $ as above,  the proportional mean  is $\alpha \approx 0.33$ with error factor $3^{\pm 1}$. With 
$\frac{z_{con}-1}{z_{con}} \approx 1$ for our curvature values $\zeta $, we  arrive at  the estimation
\[   \epsilon _{back} = \alpha ^{-1} \epsilon _{IOUV}\frac{z_{con}-1}{z_{con}} \approx 0.72 \cdot 6^{\pm 1} \; eV cm^{-3} = 7.2 \cdot 6^{\pm 1} \cdot 10^{-4} \; keV cm^{-3} .\]
The corresponding Planck  temperature is
\[   T_{back}  = k_B^{-1} \left( \frac{\epsilon _{back}}{a_0}   \right )^{\frac{1}{4}} , \;  \;  \; a_0 = \frac{\pi ^2}{15 ( \hbar c)^3} ,  \] 
  ($k_B$ Boltzmann constant).  We arrive at the conclusion:

{\em From the actual best values for the infrared, optical and UV extragagalactic  radiation} \cite{Dwek/Hauser:2001} we get an estimation for the temperature of the Segal-Olbers-Weyl background of
\[ T_{back} \approx 3.3 \cdot 1,6^{\pm 1}\, °K, \;\; \mbox{i.e.} \;\; T_{back} \in [2.1 , 5.3] \, ° K .\]

This is in encouraging agreement with the empirically determined radiation temperature of the cosmic microwave background $T_{CMB} \approx 2.73° K$.

Moreover, we observe an interesting behaviour of angular sizes in Einstein-Weyl universes: For small redshifts they decrease rapidly, but over  a long $z$- interval they stay relatively stable, before they rise again close to the conjugate point.  For $\zeta =0.8 $, the inhomogeneity scale of superclusters of about $ 50 \, Mpc$ leaves optical signals  of angular size between 0.6° and 0.8° in  87 \% of the volume of the spatial section up to the first conjugate point.

\begin{figure}[here]
\center{\includegraphics*[height=6.5cm,width=10cm]{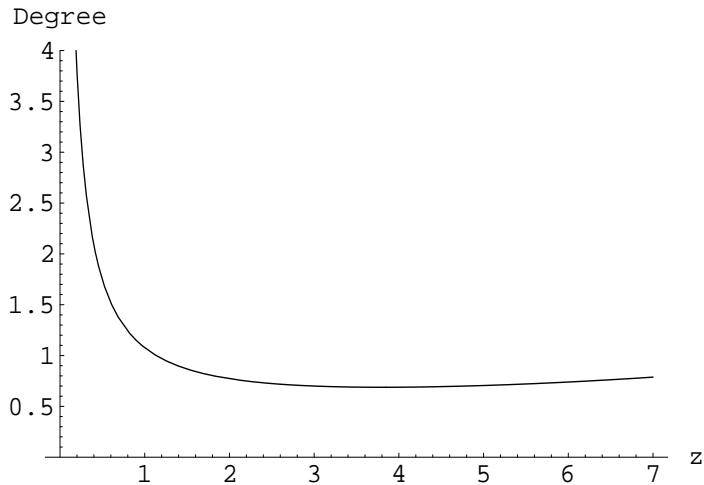}}
\caption{Angular size of objects with diameter 50 Mpc in Einstein-Weyl universe with $\zeta = 0.8$}
\end{figure}

Under the plausible assumption that anisotropies in the energy density of the  foreground field lead to fluctuations of the background  temperature  in comparable angular sizes, superclusters will cause   anisotropy traces on  the Segal-Olbers background betwen  0.6° and 0.8°  for $\zeta =0.8$. We thus expect an anisotropy signal with  multipole moment about $ l \approx 180° / 0.8 ° \approx  220$ and find  an acceptable agreement with the fine structure measurements of CMB anisotropies with a strong maximum close to $l \approx 200$ \cite{Mauskopfea}. 

Of course, this observation is far from a ``prediction'' of anisotropy angles. The determination of parameters is still much too rough to allow reliable  conclusions in this regime. If 
 $\zeta =0.3$  turns out to be  a more realistic curvature parameter of the Weyl universe, we even have to consider the inhomogeneity scale of cellular structures, of about $ 100\, Mpc$ as possibly lying at the origin of the dominant anisotropy  moment. In a volume of more than 70 \% of the spatial section they lead here to   anisotropies of  angular size between 0.75° and 1°.\\[7mm]

\subsection*{Provisional conclusion  }
The Weyl geometric approach to cosmology seems to offer great opportunities. In a wide interval of curvature parameters,  $\zeta \in [-0.4, 2.2]$, Weyl universes are in good agreement with supernovae Ia measurements. A fair interval  of Einstein-Weyl universes, $\zeta \in  [\, 0.25, 1\,]$,  is consistent with quasar data and estimations of dynamical mass for the more generous estimations on  larger scales. The models constructed in the Weyl approach  behave more trustworthy than the ones from the Friedman-Robertson  class in several respects. In particular, they are not subject to the anomaly of a time-dependent  vacuum energy  contribution, which irritates  the standard approach \cite{Carroll:Constant}.  The densities of mass-energy  and of vacuum energy are constant, considered in the mean over large  cosmic regions. Of course, the physical explanation of $\rho_{\Lambda }$ and  $p_{\Lambda }$ remains a puzzle. Its answer is no question of geometry but of physics, perhaps even of astronomy. If the contribution of the intergalactic diluted (ordinary) mass turns out to be higher than estimated at the moment, the necessity to 
bring ``vacuum energy'' (and vacuum pressure) into  play vanishes.
 But already now,  the whole confused and confusing debate about ``non-baryonic'' dark matter and time-dependent ``vacuum'' energy  becomes  superfluous, if we accept the Weyl approach as an alternative to the present standard. 

We only have to draw the natural conclusion from the differences between the values of dynamical mass concluded from astronomical observations and the prediction from the 
hypothesis of primordial nucleosyntheis. Such differences started already to become apparent  about  a decade  ago. The development in the meantime shows that it  is time to give up the reality claim of the hypothesis of {\em primordial} nucleosynthesis. 

Of course, this  long overdue conclusion  is much easier to  draw  in a reflection of the state of the art from the point of view of the Weyl approach. In this perspective,   the  whole game with high energy states close to the big bang  looks very much   like  grand talk about   a  gauge effect. The Weyl approach makes it very likely that ``physics close to the big bang'' is a pure  theory-construct, without direct physical meaning. It may be acceptable, however, as a challenging  theoretical test ground  for extreme conditions. Perhaps it continues to be useful in this function, if pursued with  the necessary precautions. For example, some of the methods developed with  the hypothesis of primordial nucleosynthesis in mind can probably   be  transferred  to the study of  hot cores of quasars and galactic nuclei and  may turn out to be useful in the new context. In the vicinity of such regional singularities of the classical geometric structure (``black holes''), which come more and more into the range of detailed  empirical investigations \cite{Peterson:Quasars},  we may expect a more realistic place for the recycling of higher nuclei to light ones than in the famous few seconds after the ``origin of the world''.

 It may be of interest to astronomers that, at the  highest redshift observations of today, differences beween the standard approach and the Weyl one start to matter and come into the range of physically relevant distinctions. For example,  the galaxy at $z \approx 7$, which  has   recently been discovered  and was precisely observed by  teams from the Keck and Hubble telescopes, appears surprisingly small in the evaluation of the cosmological standard model (with $(\Omega_m,\Omega _{\Lambda }) = (0.3, 0.7)$). The galaxy seems to have  major and minor axes of   $1.6 \,kp$ and $0.5\, kpc$ only   \cite{Kneib:Galaxy7}. Because of the high luminosity of the galaxy ($M  \approx  -21$), it has been perceived as extremely   compact and seems to be characterized by  a particularly high rate of star formation. These properties were already tentatively interpreted as an indicator for  evolutionary ``early'' galaxies. If considered in the Weyl approach,  the $z \approx 7$ galaxy comes  much  closer to normality.  In a Weyl universe of $\zeta \approx 0.8$ the axes would be calculated as  $ 19 \, kpc $ and $ 5 \, kpc  $,  in a Weyl universe with  $\zeta \approx 0.3$  even as $ 29 \, kpc $ and $ 8 \, kpc  $. 

All in all, the outcome of the first exploration of the Weyl geometric approach to cosmology is a promising first step. It cannot be more.   The next step should be a closer evaluation of the consequences of the approach for astronomical and astrophysical data, done with much more expertise in the diverse fields than stands to our disposal. On the other hand, diverse theoretical questions are open for investigation; most important, perhaps, the one for a physical explanation of the Hubble form. This component of the Weylian structure seems to indicate  a hitherto neglected coupling between gravitation and the electromagnetic field, expressed  on the classical level. The coupling is definitely different from the one proposed by   Weyl in his first hypothesis of 1918; it   rather  may have   some affinity to F. Zwicky's   proposal  for an explanation of the cosmological redshift by a kind of ``gravitational drag'' of photons \cite{Zwicky:Redshift}. H. Weyl, at least, considered this attempt as a legitimate  alternative to the space kinematical explanations of Friedman-Robertson type, even at a time when he was well aware of the necessity to find  quantum theoretical versions for the successful classical field theories \cite{Weyl:Redshift}.

As in any new approach, we should not be too sure about empirical reliability in the long run.
It may still turn out that  broader and more detailed scrutinizing of the properties of Weyl geometrical models  leads to the conclusion  that   this approach suggests  a {\em ``second}  physically wrong'', i.e., empirically inadequate, theory of  cosmic geometry.\footnote{W. Lenz, Hamburg, used to warn that for all complexes of empirical phenomena there exist ``{\em at least two} theories which are mathematically consistent and fit the data, but are physically wrong '' (communication due to W. Kunz at the Jordan Symposium, October 2004, Mainz).}
At present, and hopefully for the foreseeable future, the approach seems to be able to  contribute to our conceptual  understanding of geometry on the level  of cosmic distances in an enlightening  manner. 



\end{document}